\documentclass[a4paper,10pt]{article}
\usepackage[cp1251]{inputenc}
\usepackage[english]{babel}
\usepackage{amsmath}
\usepackage{amssymb}
\usepackage{graphicx}
\usepackage[pdftex]{color}
\usepackage[square,numbers,comma,sort&compress]{natbib}
\usepackage[pdftex]{hyperref}
\hypersetup{
             final,
             colorlinks=true,
             linkcolor=blue,
             citecolor=red
}

\begin{document}
\twocolumn[\scriptsize{\slshape ISSN 0021-3640, JETP Letters, 2014,
Vol. 100, No. 5, pp. 311--318. \textcopyright\, Pleiades Publishing,
Ltd., 2014.}

\scriptsize{\slshape Original Russian Text \textcopyright\, P.V.
Ratnikov, A.P. Silin, 2014, published in Pis'ma v Zhurnal
Eksperimental'noi i Teoreticheskoi Fiziki, 2014, Vol.
100, No. 5, pp. 349--356.}

\vspace{0.5cm}

\hrule\vspace{0.07cm}

\hrule

\vspace{0.75cm}

\begin{center}
\LARGE{\bf Novel Type of Superlattices Based on Gapless Graphene}

\LARGE{\bf with the Alternating Fermi Velocity}

\vspace{0.25cm}

\large{\bf P. V. Ratnikov and A. P. Silin}

\vspace{0.1cm}

\normalsize

\textit{Lebedev Physical Institute, Russian Academy of Sciences, Moscow, 119991 Russia}

\textit{e-mail: ratnikov@lpi.ru}

Received July 1, 2014
\end{center}

\vspace{0.1cm}
\begin{list}{}
{\rightmargin=1cm \leftmargin=1cm}
\item
\small{We study a novel type of graphene-based superlattices formed owing to a periodic modulation of the Fermi surface. Such a modulation is possible for graphene deposited on a striped substrate made of materials with
substantially different values of the dc permittivity. Similar superlattices appear also in graphene sheets
applied over substrates with a periodic array of parallel grooves. We suggest a model describing such superlattices. Using the transfer-matrix technique, we determine the dispersion relation and calculate the energy spectrum of these superlattices. We also analyze at a qualitative level the current--voltage characteristics of the system under study.}

%

\vspace{0.05cm}

\small{\bf DOI}: 10.1134/S0021364014170123

\end{list}\vspace{1.25cm}]

\begin{center}
1. INTRODUCTION
\end{center}

Graphene-based superlattices still attract considerable current interest. In particular, researchers deal
with rippled graphene, which can be considered as a superlattice with a one-dimensional periodic potential related to these ripples \citep{Isacsson}. The theoretical analysis was performed for superlattices arising under the effect of the applied periodic electrostatic potential \citep{Bai} as well as of periodic arrays formed by magnetic barriers \citep{Masir}.

\vspace{0.15cm}

Graphene-based superlattices with a periodically modulated band gap were studied in \citep{Ratnikov1}. This modulation is possible due to the interaction of graphene with the substrate material. Hexagonal boron nitride was chosen as such a material.

\vspace{0.15cm}

A similarity between the one-dimensional graphene-based superlattice and graphene exposed to
a standing laser-produced light wave was demonstrated in \citep{Savelev}. The \emph{ab initio} calculations of the electronic properties of superlattices formed by alternating of gapless graphene and graphane were reported in \citep{Lee}.

\vspace{0.15cm}

The Bloch-Zener oscillations in the superlattices formed by graphene and gapless HgTe semiconductor
were studied in \citep{Krueckl}. The splitting of the Landau levels in the graphene-based superlattice in a magnetic field applied perpendicularly to its surface was analyzed in \citep{Pal}. Different types of semiconductor superstructures were described in detail in review \citep{Sorokin}.

\vspace{0.15cm}

The model of superlattice suggested in \citep{Ratnikov1} was used in \citep{Maksimova} to find the condition corresponding to the formation of additional Dirac points and Tamm minibands arising due to the intersection of the dispersion curves of the gapless graphene and its gapped modification. This condition is in agreement with that obtained earlier for semiconductor heterostructures \citep{Kolesnikov1, Andryushin}. The surface states in graphene-based heterostructures were analyzed in \citep{Ratnikov2, Ratnikov3}.

\vspace{0.15cm}

In this paper, we suggest a novel type of superlattices based on gapless graphene with alternating
regions characterized by different values of the Fermi velocity. In our case, the \emph{Fermi velocity engineering} is based on the usage of the surrounding graphene materials, which have different values of permittivity \citep{Hwang}. The idea to control the Coulomb interaction between charge carriers in graphene by the choice of substrate materials with the necessary values of dc permittivity was first put forward in \citep{Ratnikov4}.

\vspace{0.15cm}

In such heterostructures, it is possible to achieve the energy quantization for charge carriers even in the
absence of potential barriers (regions with wider band gaps) and quantum wells (regions with narrower band
gaps), and even without any variations in the work function \citep{Kolesnikov2}. Note that low-energy (Tamm) minibands are absent here since the straight dispersion lines do not intersect anywhere except for the Dirac point.

\vspace{0.15cm}

Such structure can be produced by the deposition of graphene on striped substrates where either the
composition (parameter $x$ in an alloy) of SiO$_{2-x}$, or the density of some (nonmagnetic) impurities, or dc
permittivity $\varepsilon$ exhibit periodic variations. Here, we treat in detail the latter possibility.

\vspace{0.15cm}

According to the results of the theoretical \citep{Gonzalez1, Gonzalez2, Sarma, Aleiner, Juan}
and experimental \citep{Hwang, Bostwick, Li1, Li2, Elias, Chae} studies, the Fermi
velocity becomes substantially renormalized. To estimate the renormalized Fermi velocity, we can use the
relation \citep{Sarma}
\begin{equation*}
\frac{\texttt{v}_\texttt{F}}{\texttt{v}_{\texttt{F}0}}\,=\,1\,-\,3.28\alpha^*\left[1\,+\,\frac{1}{4}\ln\left(1\,+\,\frac{1}{4\alpha^*}\,-\,1.45\right)\right],
\end{equation*}
where $\alpha^*\,=\,\widetilde{e}^2/\hbar\texttt{v}_{\texttt{F}0}$ is the analog of the fine structure
constant, $\texttt{v}_{\texttt{F}0}$ is the initial unrenormalized Fermi velo-city ($\texttt{v}_{\texttt{F}0}\,=\,0.85\times10^8$ cm/s) \citep{Hwang, Elias}, $\widetilde{e}^2\,=\,e^2/\varepsilon_{eff}$, and $\varepsilon_{eff}\,=\,(\varepsilon_1\,+\,\varepsilon_2)/2$ is the effective dc permittivity for the charge carriers in graphene depending on the values $\varepsilon_1$ and $\varepsilon_2$ of dc permittivity characterizing the materials surrounding graphene. Note that here the band gap is not open; this is confirmed in experiment with an accuracy of 0.1 meV \citep{Elias}.

\begin{figure}[t!]
\begin{center}
\hypertarget{fig1}{}
\hypertarget{fig1a}{}
\includegraphics[width=0.5\textwidth]{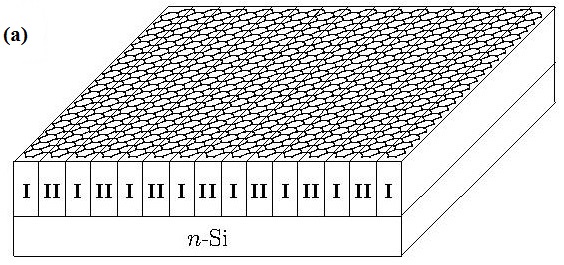}

\vspace{0.2cm}

\hypertarget{fig1b}{}
\includegraphics[width=0.5\textwidth]{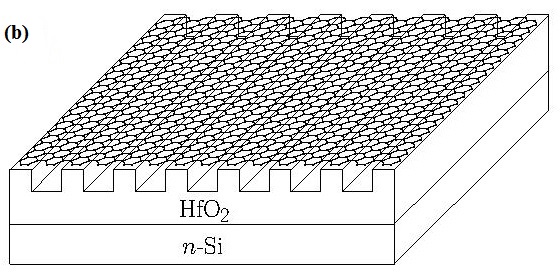}

\vspace{0.2cm}

\hypertarget{fig1c}{}
\includegraphics[width=0.5\textwidth]{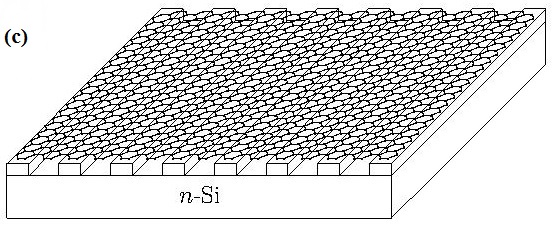}
\end{center}

\small{{\bf Fig. 1.} Three variants of the superlattice under study: {\bf(a)}~graphene sheet placed on a striped substrate consisting of alternating layers of materials with substantially different
values of the permittivity, e.g., SiO$_2$ with $\varepsilon\,=\,3.9$ (I) and
HfO$_2$ with $\varepsilon\,=\,25$ (II); {\bf(b)} graphene sheet placed on the
HfO$_2$ substrate with periodically arranged grooves; and {\bf(c)}~graphene sheet deposited on a periodic array of parallel metallic strips. A plate of heavily doped silicon \emph{n}-Si is used
as a gate.}
\end{figure}

\vspace{0.15cm}

Within the graphene region located over the strip with the lower value of $\varepsilon$, we have larger $\alpha^*$. Hence, the corresponding renormalized Fermi velocity should be higher than that over the strip with the higher value of $\varepsilon$. This suggests the possibility of modulating $\texttt{v}_\texttt{F}$ by varying the substrate permittivity. Note that such a system is a one-dimensional photonic crystal.

\vspace{0.15cm}

The first version of the suggested system is a gra-phene sheet placed on a striped substrate consisting
of alternating layers of materials with substantially different values of the permittivity. A schematic image of such a system is shown in \hyperlink{fig1a}{Fig. 1a}.

\vspace{0.15cm}

It is also possible to use a substrate with periodically arranged grooves prepared by etching. The graphene
sheet placed on such substrate should have the periodically alternating regions suspended over the grooves
and those being in contact with the substrate material (see \hyperlink{fig1b}{Fig. 1b}). The renormalization of the Fermi velocity should be the most clearly pronounced just in the suspended graphene regions since here we have $\varepsilon_{eff}\,=\,1$. According to the experimental data, the renormalized Fermi velocity in suspended graphene increases to $3\times10^8$ cm/s \citep{Elias}.

\vspace{0.15cm}

In the regions with graphene in contact with the narrow gap semiconducting material, where $\varepsilon_{eff}\,\gg\,1$, the renormalized Fermi velocity differs only slightly from the unrenormalized one. In addition, the substrate itself is a diffraction grating. Therefore, the system should exhibit rather interesting optical characteristics, demanding a separate study.

\vspace{0.15cm}

There is another version of the system under study. It is possible to deposit graphene on a periodic array of
parallel metallic strips (\hyperlink{fig1c}{Fig. 1c}). This is the limiting case: in the suspended graphene regions, we have $\varepsilon_{eff}\,=\,1$ (the strongest renormalization of the Fermi velocity), whereas in the regions with graphene in contact with metallic strips, we have $\varepsilon_{eff}\,=\,\infty$ (vanishing renormalization of the Fermi velocity \citep{Hwang}).

\vspace{0.15cm}

We see that a whole class of such type of systems, which were not studied earlier, is possible. Without
doubt, the studies of such systems should lead to important advances in the implementation of the technologies based on the controlled Fermi velocity.

\vspace{0.15cm}

\begin{center}
2. MODEL
\end{center}

The model for the description of the suggested superlattice is similar to that used earlier to study the
superlattice on the striped substrate with the periodic variation in the band gap \citep{Ratnikov1}.

\vspace{0.15cm}

In our case, we assume that the band gap remains unchanged and is equal to zero (gapless graphene) and
the work function is the same over all regions of the superlattice (its value is chosen as the energy reference point). We have only a modulation of the Fermi velo-city. In gapless graphene, a change in the work function
leads to the electrical breakdown and to the creation of electron-hole pairs. We also assume that the near-border region corresponding to the gradual change in the Fermi velocity is much narrower than the superlattice period. Therefore, the $\texttt{v}_\texttt{F}$ profile can be considered to be sharp enough (see \hyperlink{fig2}{Fig. 2}).

\begin{figure}[t!]
\begin{center}
\hypertarget{fig2}{}
\includegraphics[width=0.5\textwidth]{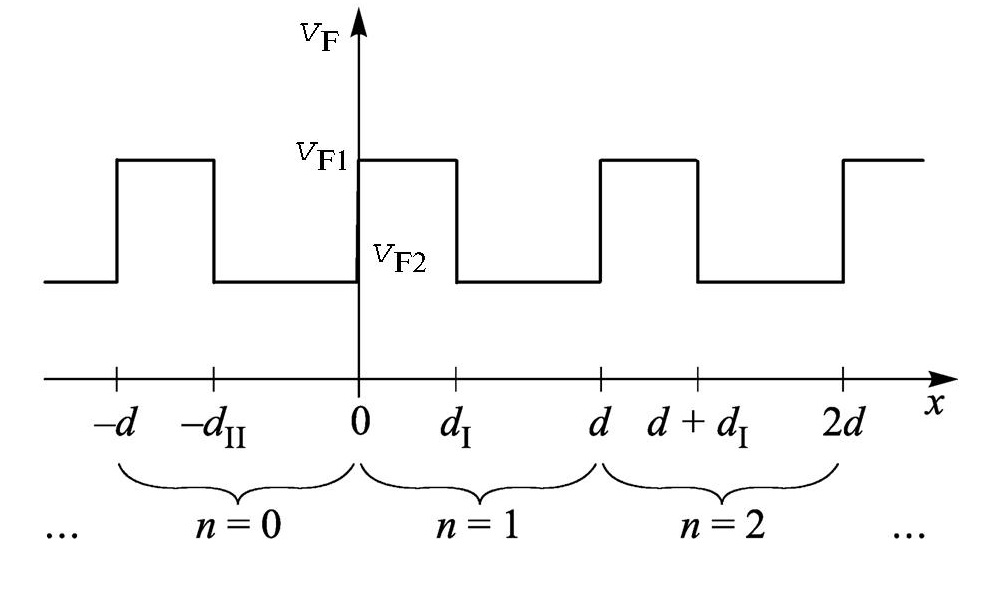}
\end{center}

\small{{\bf Fig. 2.} Fermi velocity profile in the superlattice under
study ($\texttt{v}_{\texttt{F}1}>\texttt{v}_{\text{F}2}$ case). The enumeration of supercells in
the superlattice and the sizes of its regions are indicated in the lower part of the figure: $d_I$ is the width of the graphene strip with the Fermi velocity $\texttt{v}_{\texttt{F}1}$, $d_{II}$ is the width of the
graphene strip with the Fermi velocity $\texttt{v}_{\text{F}2}$, and $d\,=\,d_I\,+\,d_{II}$ is the period of the superlattice.}
\end{figure}

\vspace{0.15cm}

We consider the charge carriers located close to the $K$ point of the Brillouin zone (the results should be the same for the charge carriers located in the vicinity of the $K^\prime$ point). Let the $x$ axis be perpendicular to the strips as is shown in \hyperlink{fig2}{Fig. 2}. The envelope of the wave-function $\Psi(x,\,y)$ for the charge carriers obeys the Dirac-Weyl equation with variable Fermi velocity\footnote{In the general case, one should write the anticommutator of the Fermi velocity $\texttt{v}_\texttt{F}(x)$ with the term containing the momentum
operator $\hat{p}_x$
\begin{equation*}
\frac{1}{2}\left\{\texttt{v}_\texttt{F}(x),\,{\boldsymbol\sigma}\hat{\bf p}\right\}\Psi(x,\,y)\,=\,E\Psi(x,\,y).
\end{equation*}
Such symmetrization of the Hamiltonian is necessary for retaining its Hermitian form. Similar problems were considered in \citep{Geller, Kolesnikov3}. In the case of the stepwise profile \eqref{2} of the Fermi
velocity, we obtain the equation for $\Psi(x,\,y)$ in form \eqref{1}. This limitation is not significant since allowance for a smooth dependence $\texttt{v}_\texttt{F}(x)$ will complicate the calculations, but will insignificantly change the final results.}
\begin{equation}\label{1}
\texttt{v}_\texttt{F}{\boldsymbol\sigma}\hat{\bf p}\Psi(x,\,y)\,=\,E\Psi(x,\,y),
\end{equation}
\begin{equation}\label{2}
\texttt{v}_\texttt{F}\,=\,\begin{cases}
\texttt{v}_{\texttt{F}1},\hspace{0.1cm}& d(n-1)\,<\,x<\,-d_{II}\,+\,dn\\
\texttt{v}_{\texttt{F}2},\hspace{0.1cm}& -d_{II}\,+\,dn\,<\,x\,<\,dn.
\end{cases}
\end{equation}
Here, $\widehat{\bf p}$\,=\,$-i\boldsymbol{\nabla}$ is the momentum operator (here and further on, $\hbar\,=\,1$). Integers $n$ enumerate supercells (see \hyperlink{fig2}{Fig. 2}). The Pauli matrices $\boldsymbol{\sigma}$\,=\,$(\sigma_x,\,\sigma_y)$ act in the space of two sublattices. The motion of charge carriers in the superlattice along the $y$ axis is free; hence, a solution to Eq. \eqref{1} has the form $\Psi(x,\,y)\,=\,\psi(x)e^{ik_yy}$.

\begin{figure}[t!]
\begin{center}
\hypertarget{fig3a}{}
\includegraphics[width=0.5\textwidth]{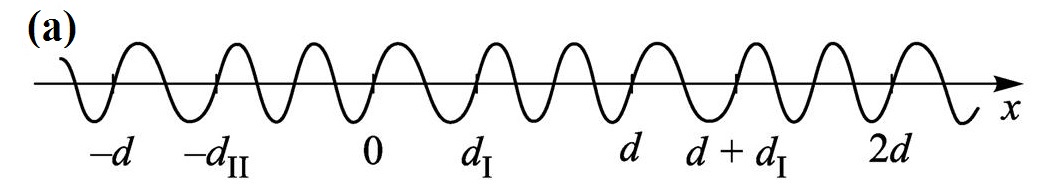}

\hypertarget{fig3b}{}
\includegraphics[width=0.5\textwidth]{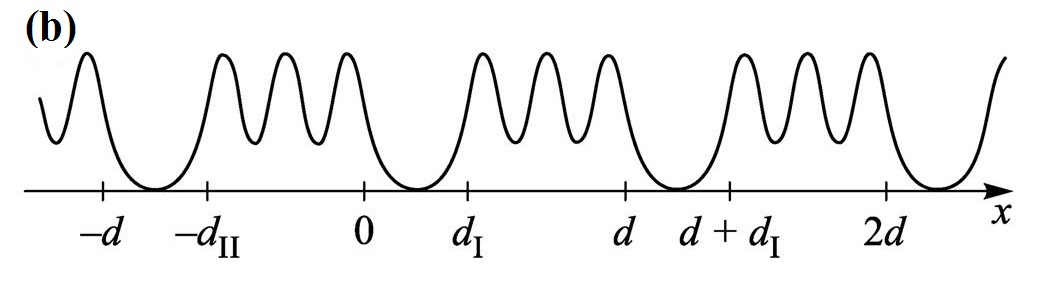}
\end{center}

\vspace{0.15cm}

\small{{\bf Fig. 3.} Schematic image illustrating the behavior of the envelope of the wavefunction of charge carriers in the superlattice under study: {\bf(a)} the oscillatory solution in all regions and {\bf(b)} the solution being oscillatory in one region and exhibiting exponential decay deep into another region
($\texttt{v}_{\texttt{F}1}>\texttt{v}_{\texttt{F}2}$ case).}
\end{figure}

\vspace{0.15cm}

Similarly to \citep{Ratnikov1}, we find a solution of Eq. \eqref{1} with
respect to $\psi(x)$ for the $n$th supercell

\vspace{0.15cm}

(i) at $0\,<\,x\,<\,d_I$
\begin{equation*}
\psi^{(1)}_n(x)\,=\,\Omega_{k_1}(x){a^{(1)}_n\choose c^{(1)}_n},
\end{equation*}
\begin{equation*}
\Omega_{k_1}(x)\,=\,N_{k_1}\begin{pmatrix}1&1\\
\lambda^{(1)}_+&-\lambda^{(1)}_-\end{pmatrix}e^{ik_1x\sigma_z},
\end{equation*}
\begin{equation*}
\lambda^{(1)}_\pm\,=\,\frac{\texttt{v}_{\texttt{F}1}(k_1\,\pm\, ik_y)}{E},\,\,\,k_1\,=\,\frac{\sqrt{E^2\,-\,\texttt{v}^2_{\texttt{F}1}k^2_y}}{\texttt{v}_{\texttt{F}1}},
\end{equation*}

\vspace{0.15cm}

(ii) at $d_I\,<\,x\,<\,d$
\begin{equation*}
\psi^{(2)}_n(x)\,=\,\Omega_{k_2}(x){a^{(2)}_n\choose c^{(2)}_n},
\end{equation*}
\begin{equation*}
\Omega_{k_2}(x)\,=\,N_{k_2}\begin{pmatrix}1&1\\
\lambda^{(2)}_+&-\lambda^{(2)}_-\end{pmatrix}e^{ik_2x\sigma_z},
\end{equation*}
\begin{equation*}
\lambda^{(2)}_\pm\,=\,\frac{\texttt{v}_{\texttt{F}2}(k_2\,\pm\,ik_y)}{E},\,k_2\,=\,\frac{\sqrt{E^2\,-\,\texttt{v}^2_{\texttt{F}2}k^2_y}}{\texttt{v}_{\texttt{F}2}}.
\end{equation*}
Here, $N_{k_1}$ and $N_{k_2}$ are the normalization factors.

\vspace{0.15cm}

For the case $\texttt{v}_{\texttt{F}1}>\texttt{v}_{\texttt{F}2}$, the condition for the existence of the solution of Eq. \eqref{1}, which oscillates in all regions of the superlattice (it is schematically illustrated in \hyperlink{fig3a}{Fig. 3a}), is reduced to the inequality
\begin{equation}\label{3}
k^2_2\,>\,\left(\frac{\texttt{v}^2_{\texttt{F}1}}{\texttt{v}^2_{\texttt{F}2}}\,-\,1\right)k^2_y.
\end{equation}

\vspace{0.15cm}

The existence of a \emph{solution of the mixed type} is also possible (see \hyperlink{fig3b}{Fig. 3b}). In this case, we have an oscillatory solution in some regions (effective quantum wells), whereas in the other regions, it exhibits exponential decay (effective potential barriers) deep into these regions. The condition for the existence of the mixed type solution is determined by the inequality inverse to \eqref{3} and it is met only for finite $k_y$ values.

\vspace{0.15cm}

The effective quantum barrier of the new type is the region with the higher Fermi velocity because the
energy of the charge carriers with the same momentum \textbf{k} in it is higher than that in the effective quantum well with the lower Fermi velocity \citep{Kolesnikov2}. In~contrast~to the usual quantum well, which is formed owing to the change in the width of the band gap, the height of the barrier in the superlattice under study grows with $k_y$. At $k_y$\,=\,0, the barrier vanishes and our problem is reduced to the \emph{empty lattice model} \citep{Callaway}. In the latter model, the potential is absent, but the periodicity is retained. As a result, energy bands corresponding to the symmetry of the problem arise, but we have zero band gaps.

\begin{center}
3. DISPERSION RELATIONS
\end{center}

To derive the dispersion relations, we use the transfer matrix ($T$-matrix) method in the way similar to that
employed in \citep{Ratnikov1}.

\vspace{0.15cm}

The transfer matrix determines the relation between the coefficients appearing in the expressions
for the envelopes of the wavefunctions for the neighboring supercells
\begin{equation*}
{a^{(1)}_{n+1}\choose c^{(1)}_{n+1}}\,=\,T{a^{(1)}_n\choose c^{(1)}_n},\,\,\,{a^{(2)}_{n+1}\choose c^{(2)}_{n+1}}\,=\,T{a^{(2)}_n\choose c^{(2)}_n}.
\end{equation*}
We use the following boundary conditions for matching of the envelopes of the wavefunctions \citep{Ratnikov3, Silin1},
\begin{equation*}
\sqrt{\text{v}_\text{F1}}\psi^{(1)}_n\,=\,\sqrt{\text{v}_\text{F2}}\psi^{(2)}_n,
\end{equation*}
and also the Bloch conditions in the form
\begin{equation*}
\psi^{(1)}_n(x+d)\,=\,\psi^{(1)}_n(x)e^{ik_xd}
\end{equation*}
and
\begin{equation*}
\psi^{(2)}_n(x+d)\,=\,\psi^{(2)}_n(x)e^{ik_xd}.
\end{equation*}
Then, the expression for the $T$-matrix has the form \citep{Ratnikov1}
\begin{equation*}
T\,=\,\Omega^{-1}_{k_1}(0)\Omega_{k_2}(d)\Omega^{-1}_{k_2}(d_I)\Omega_{k_1}(d_I).
\end{equation*}
The dispersion relation is determined from the equality
\begin{equation*}
\text{Tr}T\,=\,2\cos(k_xd),
\end{equation*}
which for the oscillatory type solution, can be written as
\begin{equation}\label{4}
\begin{split}
\frac{\texttt{v}_{\texttt{F}1}\texttt{v}_{\texttt{F}2}k^2_y-E^2}{\texttt{v}_{\texttt{F}1}\texttt{v}_{\texttt{F}2}k_1k_2}\sin(k_1d_I)\sin(k_2d_{II})\\
+\,\cos(k_1d_I)\cos(k_2d_{II})\,=\,\cos(k_xd).
\end{split}
\end{equation}
For the solution of the mixed type, the dispersion relation is found from \eqref{4} through the use of the formal substitution $k_1\,\rightarrow\,i\kappa_1$, where $\kappa_1\,=\,\frac{1}{\texttt{v}_{\texttt{F}1}}\sqrt{\texttt{v}^2_{\texttt{F}1}k^2_y\,-\,E^2}$.

\vspace{0.15cm}

At $k_y\,=\,0$, transcendental equation \eqref{4} has the form
\begin{equation}\label{5}
\cos\left(k_1d_I\,+\,k_2d_{II}\right)\,=\,\cos(k_xd)
\end{equation}
for which the exact solution can be found
\begin{equation*}
E_l(k_x)\,=\,\pm\texttt{v}^*_\texttt{F}\left(k_x\,+\,\frac{2\pi l}{d}\right), \,\,\,l=0,\,1,\,2,\,\ldots\,.
\end{equation*}
Here, the effective Fermi velocity is introduced as
\begin{equation}\label{6}
\texttt{v}^*_\texttt{F}\,=\,\frac{\texttt{v}_{\texttt{F}1}\texttt{v}_{\texttt{F}2}d}{\texttt{v}_{\texttt{F}1}d_{II}+\texttt{v}_{\texttt{F}2}d_I}.
\end{equation}

\vspace{0.15cm}

For the $l$th miniband, the energy at the $K$ point is equal to
\begin{equation*}
E^0_l\,=\,\pm\frac{2\pi l\texttt{v}^*_\texttt{F}}{d},\,\,\,l=0,\,1,\,2,\,\ldots\,.
\end{equation*}
We can see that the lower electron miniband ($l\,=\,0$) touches the upper hole miniband at the $K$ point and graphene remains gapless.

\vspace{0.15cm}

From Eq. \eqref{5}, we find that, at the edge of the $l$th miniband, the energy at $k_x\,=\,\pm\pi/d$ is equal to
\begin{equation*}
E_l\left(\pm\frac{\pi}{d}\right)\,=\,\pm\frac{\pi(2l+1)\texttt{v}^*_\texttt{F}}{d},\,\,\,l=0,\,1,\,2,\,\ldots\,.
\end{equation*}
The minibands are separated by the direct band gaps
\begin{equation*}
E_G\,=\,E_{l+1}\left(\pm\frac{\pi}{d}\right)\,-\,E_l\left(\pm\frac{\pi}{d}\right)\,=\,\frac{2\pi\texttt{v}^*_\texttt{F}}{d}.
\end{equation*}
In the case of $k_y\,=\,0$, indirect gaps are absent
\begin{equation*}
E_l\left(\frac{\pi}{d}\right)\,=\,E_{l+1}\left(-\frac{\pi}{d}\right),
\end{equation*}
which corresponds to the empty lattice model \citep{Callaway}.

\vspace{0.15cm}

\begin{center}
4. NUMERICAL CALCULATION\\
OF THE ENERGY SPECTRUM
\end{center}

Let us calculate the lower electron miniband for the superlattice shown in \hyperlink{fig1c}{Fig. 1c}. According to \citep{Elias}, for it we have $\texttt{v}_{\texttt{F}1}\,=\,3\times10^6$ cm/s (suspended graphene) and $\texttt{v}_{\texttt{F}2}\,=\,0.85\times10^6$ cm/s (in the region with the contact of graphene with the metal, the Fermi velocity coincides with $\texttt{v}_{\texttt{F}0}$).

\vspace{0.15cm}

In the weak coupling model, the problem concerning the edge type at the interface turns out to be unimportant. Let us assume that we have a zigzag-type boundary at the interface (see \hyperlink{fig1}{Fig. 1}) and, in each of two regions of the supercell, integer numbers $N_I$ and $N_{II}$ of graphene unit cells are packed up. Then, we have $d_I\,=\,3N_Ia$ and $d_{II}\,=\,3N_{II}a$, where $a\,=\,1.42$~\AA\, is the lattice constant of graphene. For calculations, we assume that $N_I\,=\,N_{II}\,=\,50$, i.e., $d_I\,=\,d_{II}\,=\,21.3$~nm.

\vspace{0.15cm}

In the framework of the suggested model, it is ne-cessary to introduce the upper limit on the wave vector
component characterizing the free motion of charge carriers, $|k_y|\ll k_c$. Momentum $k_c$ corresponds to the
energy of the ultraviolet cutoff, $\Lambda\,\approx\,3$ eV \citep{Elias}. As a result, we find $k_c\,\approx\,4.3$ nm$^{-1}$. This, in turn, imposes the limitation on the superlattice period, $d\gg a$.

\vspace{0.15cm}

The results of numerical calculations are represented in the form of two $E(k_x,\,k_y)$ plots for the lower
electron miniband: (i) $E(k_x)$ at fixed values of $k_y$ (\hyperlink{fig4}{Fig. 4a}) and (ii) $E(k_y)$ at fixed values of $k_x$ (\hyperlink{fig4}{Fig. 4b}). In \hyperlink{fig4}{Fig.~4a}, we can see, in particular, that $k_y\,=\,0$ corresponds to the linear dispersion law and the effective Fermi velocity is $\texttt{v}^*_\texttt{F}\,\approx\,1.325\times10^8$ cm/s. The lower curve in \hyperlink{fig4}{Fig. 4b} exhibits a nearly linear growth. This means that the $E(k_x,\,k_y)$ surface has the conical shape near the Dirac point.

\vspace{0.15cm}

Thus, we confirm by numerical calculations that at $k_y\,=\,0$, the Fermi velocity of electrons (holes) has a
constant value, does not vanish up to the boundaries of minibands, and is determined by Eq. \eqref{6} (this is true for all minibands). In this sense, the particles do not feel the boundaries of minibands. Note that, for $k_y\,\neq\,0$, the velocity of particles always vanishes at the miniband boundaries.

\begin{figure}[t!]
\begin{center}
\hypertarget{fig4}{}
\includegraphics[width=0.5\textwidth]{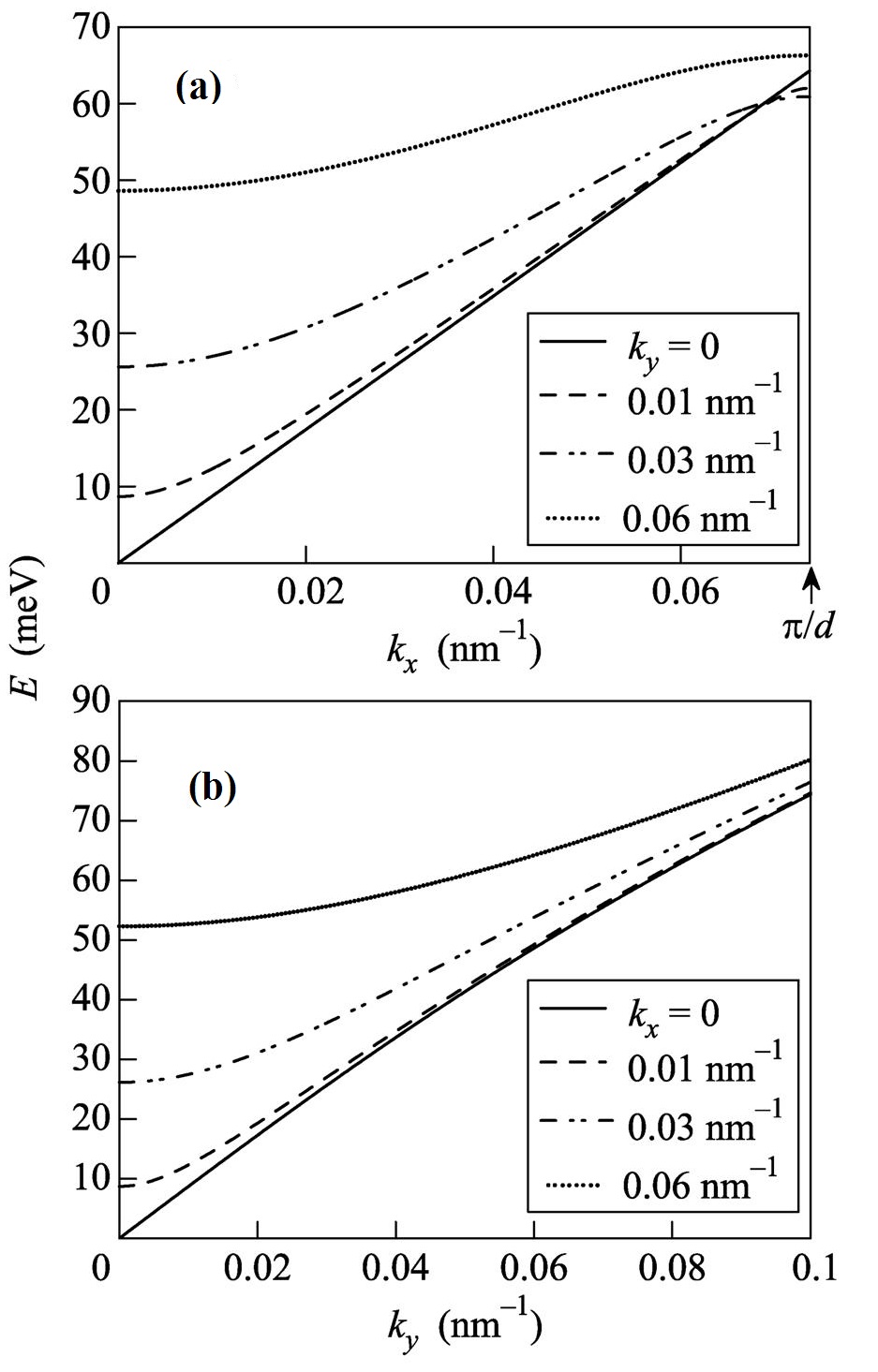}
\end{center}

\small{{\bf Fig. 4.} Numerical calculation of the dispersion curves for the lower electron miniband {\bf(a)} versus $k_x$ at fixed $k_y$ values and {\bf(b)} versus $k_y$ at fixed $k_x$ values.}
\end{figure}

\vspace{0.5cm}
\begin{center}
5. CURRENT-VOLTAGE CHARACTERISTICS\\
(QUALITATIVE ANALYSIS)
\end{center}

\vspace{0.3cm}

Let us briefly discuss at the qualitative level the effect of the superlattice potential on the transport
phenomena.

\vspace{0.27cm}

Having in mind the aforementioned qualitative difference between the $k_y = 0$ and $k_y \neq 0$ cases, we should expect that the current-voltage characteristics ($I$--$V$ curves) of the superlattice under study should be significantly different for these two cases.

\vspace{0.27cm}

At $k_y\,=\,0$, the transport characteristics of the superlattice under study should be the same as for effective gapless graphene with the average Fermi velocity $\texttt{v}^*_\texttt{F}$ given by Eq. \eqref{6}. In particular, at any arbitrarily low charge carrier density, we should observe nonzero minimum conductivity $\sigma_{min}$. According to the experimental data, we have $\sigma_{min}\,=\,4e^2/h$ \citep{Novoselov}, which coincides with the ballistic conductivity of graphene. The $I$--$V$ curve should exhibit a linear growth similar to that characteristic of graphene samples with high enough mobility of charge carriers, $\mu\gtrsim10^4$ cm$^2$/(V\,s) \cite{Vandecasteele}.

\vspace{0.15cm}

In the case of $k_y\,\neq\,0$, the situation is more complicated. At a nonzero transverse field $V_y$ and at a sufficiently small longitudinal field $V_x$, the $I$--$V$ curve should be a growing one and the differential conductivity at small values of $V_x$ is about or higher than the minimum conductivity
\begin{equation*}
\sigma_{dif}(V_x\approx0)\gtrsim\sigma_{min}.
\end{equation*}

\vspace{0.15cm}

Now, we calculate the velocity of electrons for the case of fixed longitudinal ($\mathcal{E}_x$) and nonzero transverse ($\mathcal{E}_y$) electric fields. For the corresponding implementation of such situation in experiment, it is possible to use the standard Hall layout.

\vspace{0.15cm}

For simplicity, we assume that transport is ballistic; i.e., the mean free path $\lambda$ is so large that an electron accelerated by the applied electric field can reach the miniband boundary without any scattering. To distinguish the spectrum related to the potential of the superlattice, the mean free path should be much larger than the period of the superlattice \citep{Silin2}
\begin{equation}\label{7}
\lambda\gg d.
\end{equation}

\vspace{0.15cm}

For the sufficiently pure graphene samples, we have $\lambda\,\simeq\,1\,\mu$m \citep{Bolotin}.

\begin{figure}[t!]
\begin{center}
\hypertarget{fig5}{}
\includegraphics[width=0.5\textwidth]{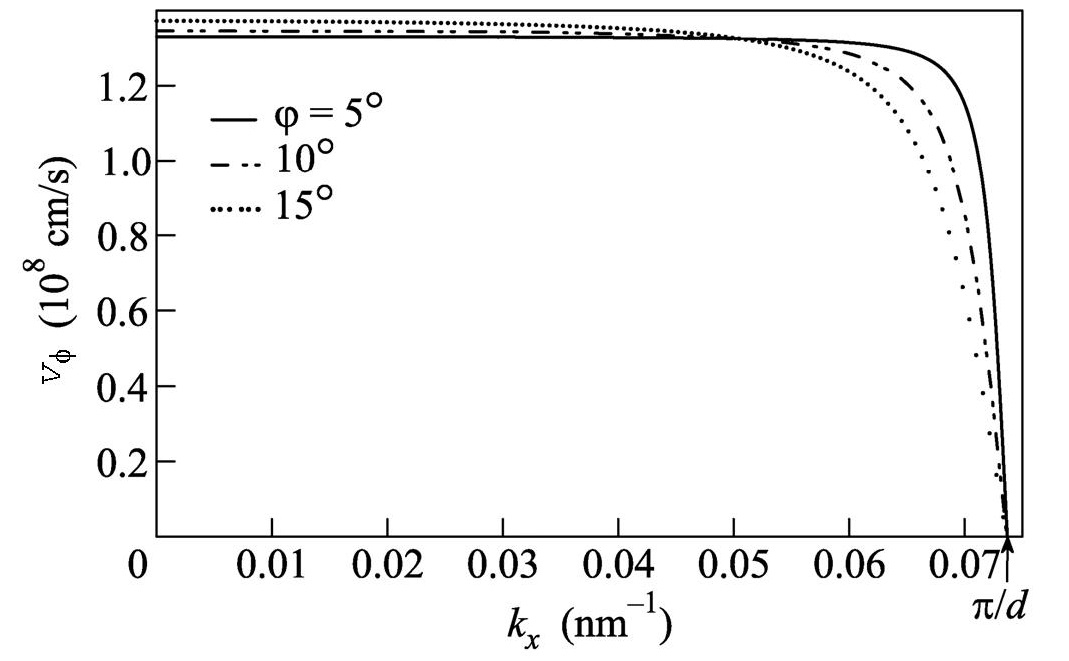}
\end{center}

\small{{\bf Fig. 5.} Numerical calculation of the electron velocity in
the lower miniband along the direction specified by the
fixed polar angle $\phi$.}
\end{figure}

\vspace{0.15cm}

The direction of the electron motion is characte-rized by the polar angle $\phi\,=\,\arctan(k_y/k_x)$. Its value remains unchanged in the whole $-\pi/d\,\leq\,k_x\,\leq\,\pi/d$ range. The contribution to the conductivity related to the intraminiband transitions is determined by the electron velocity, which we seek:
\begin{equation*}
\texttt{v}_\phi\,=\,\left.\frac{\partial E}{\partial k}\right|_{k_y=k_x\tan\phi}.
\end{equation*}

\vspace{0.15cm}

In \hyperlink{fig5}{Fig. 5}, we illustrate the calculated dependence of the electron velocity on $k_x$ for the same superlattice parameters as above for the polar angles $\phi\,=\,5^o, 10^o,$
and $15^o$. We can see that the velocity indeed vanishes at the miniband boundary and its abrupt decrease
takes place within a quite narrow range near the miniband boundary. For low momenta, we have $\texttt{v}_\phi\approx\texttt{v}^*_\texttt{F}$.

\vspace{0.15cm}

An application of the superlattice at nonzero temperatures requires the existence of a quite clearly pronounced Fermi velocity profile; i.e., we should use rather large $\phi$ and $\delta\texttt{v}_\texttt{F}\,=\,\left|\texttt{v}_{\texttt{F}1}\,-\,\texttt{v}_{\texttt{F}2}\right|$ values:
\begin{equation*}
\pi\frac{\delta\texttt{v}_\texttt{F}}{d}\sin\phi\gg T.
\end{equation*}
However, at large $\phi$ values close to $\pi/2$, the condition according to which charge carriers pass a large number of supercells at the mean free path can be violated. Then, condition \eqref{7} turns out to be unimportant (condition $l\cos\phi\gg d$ should be met).

\begin{figure}[t!]
\begin{center}
\hypertarget{fig6}{}
\includegraphics[width=0.5\textwidth]{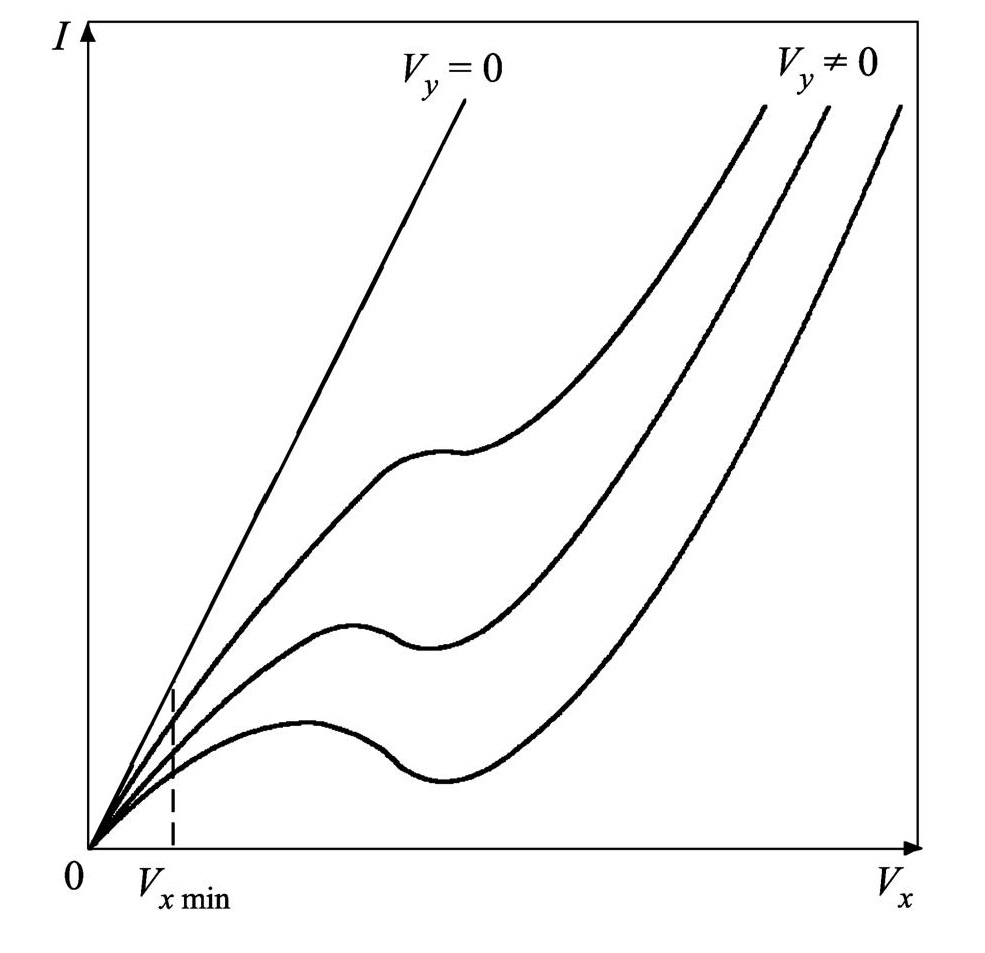}
\end{center}

\small{{\bf Fig. 6.} Qualitative behavior of the $I$--$V$ curve for the superlattice under study. Three $I(V_x)$ plots under the linear $I$--$V$ curve correspond to the growth of the transverse voltage $V_y$ (from top to bottom).}
\end{figure}

\vspace{0.15cm}

Similarly to the situation occurring in semiconductor superlattices, the motion of charge carriers at sufficiently strong electric field $\mathcal{E}_x$ is finite. They oscillate with the Stark frequency \citep{Krueckl, Silin2}
\begin{equation*}
\Omega\,=\,e\mathcal{E}_xd.
\end{equation*}
This stems from the nonlinearity of the $I$--$V$ curve manifesting itself in the negative differential conductivity at a certain section of it. Charge carriers in the nonlinear regime undergo a large number of the Bloch oscillations during the mean free time $\tau$:
\begin{equation}\label{8}
\Omega\tau\gg1.
\end{equation}
We estimate the mean free time as $\tau\approx\lambda/\texttt{v}^*_\texttt{F}$ (the velo-city of charge carriers is $\texttt{v}_\phi\approx\texttt{v}^*_\texttt{F}$ everywhere except for a narrow range near the miniband boundaries). Then, condition \eqref{8} can be rewritten as
\begin{equation}\label{9}
\mathcal{E}_x\gg\frac{\texttt{v}^*_\texttt{F}}{ed\lambda}.
\end{equation}
Condition \eqref{9} automatically gives an estimate for the minimum longitudinal voltage above which negative
differential conductivity becomes possible
\begin{equation*}
V_{x\,\min}\,\simeq\,\frac{\texttt{v}^*_\texttt{F}}{ed}\frac{L_x}{\lambda},
\end{equation*}
where $L_x$ is the size of the system along the $x$ axis. Assuming that $L_x\,\simeq\,\lambda$, we arrive at the estimate $V_{x\,min}\,\simeq\,0.02$ V for the superlattice with the same parameters as above.

\vspace{0.15cm}

In \hyperlink{fig6}{Fig. 6}, we represent the qualitative behavior of the $I$--$V$ curve for the superlattice under study. At $k_y\,=\,0$ (zero applied voltage in the transverse direction, $V_y\,=\,0$), we observe its linear growth. At $k_y\,\neq \,0$ (nonzero transverse voltage, $V_y\,\neq\,0$), a section with negative differential conductivity arises in the curve. In this case, for higher $V_y$ values, this section is more pronounced and more shifted toward lower $V_x$ values. However, as is mentioned above, this section can arise only at a sufficiently high longitudinal voltage
\begin{equation*}
V_x\gg V_{x\min}.
\end{equation*}

\vspace{0.15cm}

Note finally that the characteristics of the system under study can depend on the gate voltage $V_g$ (at different values of the charge carrier density $n_{2D}$) owing to the dependence of the renormalized Fermi velocity on $n_{2D}$ \citep{Elias, Chae}. In this case, a controlling factor is the filling of minibands with electrons (holes). For the experimental observations, it is convenient to have partially filled either the lower electronic miniband or the upper hole one (in this case, the higher electronic or lower hole minibands are distinguishable). This takes place if $n_{2D}\ll n^*_{2D}=4/d^2$. This condition can be rewritten in the form of a limitation imposed on the gate voltage
\begin{equation*}
|V_g|\ll4\pi en^*_{2D}L_g/\varepsilon^*_s,
\end{equation*}
where $L_g$ is the gate thickness and $\varepsilon^*_s$ is the effective dc
permittivity of the substrate. For the layered substrate structure (see \hyperlink{fig1a}{Fig. 1a}), we have
\vspace{0.15cm}
\begin{equation*}
\varepsilon^*_s=\frac{\varepsilon_{s1}d_I+\varepsilon_{s2}d_{II}}{d}.
\end{equation*}

\vspace{0.25cm}

\begin{center}
6. CONCLUSIONS
\end{center}

\vspace{0.25cm}

We suggested a novel class of graphene-based systems, which are at the same time both photon crystals
and graphene superlattices with periodically varying Fermi velocity. Such a modulation appears to be possible owing to the renormalization of the Fermi velocity in the energy spectrum of graphene. New prospects become open for the implementation of the technologies based on controlled Fermi velocity. We point out some specific features of the transport phenomena in such systems, in particular, appearance of the sections with negative differential conductivity in the $I$--$V$ curves. It is clear that, similarly to photon crystals,
these systems should exhibit interesting optical characteristics.

\vspace{1cm}\flushright\textit{Translated by K. Kugel}
\end{document}